\documentstyle[prl,aps,psfig]{revtex}

\begin{document}

\title{Supernova SN1987A Bound on Neutrino Spectra for R-Process Nucleosynthesis}

\author{C.~J.~Horowitz
\footnote{e-mail:  horowitz@iucf.indiana.edu} 
}
\address{Nuclear Theory Center and Dept. of Physics, Indiana
University, Bloomington, IN 47405}
\date{\today} 
\maketitle 
\begin{abstract}
The neutrino driven wind during a core collapse supernova is an attractive site for r-process nucleosynthesis.  The initial electron fraction $Y_e$ in the wind depends on observable neutrino energies and luminosities.  The mean antineutrino energy is limited by supernova SN1987A data while lepton number conservation constrains the ratio of antineutrino to neutrino luminosities.  If $Y_e$, in the wind, is to be suitable for rapid neutron capture nucleosynthesis, then the mean electron neutrino energy could be significantly lower then that predicted in present supernova simulations.  Alternatively, there could be a rapid increase of the mean antineutrino energy at late times.  However, this is not seen in the SN1987A data.  Finally, there could be new neutrino physics such as oscillations to sterile neutrinos.

\end{abstract}
\vskip .2in

The neutrino driven wind above a protoneutron star in a core collapse supernova is an attractive site for r-process nucleosynthesis.  In the r-process, seed nuclei rapidly capture free neutrons to produce about half of the heavy elements\cite{bur57,wal97}.  Many simulations, for example \cite{bra97}, have explored a range of physical conditions including entropy, expansion time scale, and electron fraction (number of electrons or protons per baryon) $Y_e$ necessary to produce r-process elements with solar system abundances.  

The inital electron fraction $Y_e$ is an important parameter that determines the number of free neutrons.  If there are too few free neutrons per seed nucleus, then the heaviest elements may not be produced.  A reasonable minimum requirement for an r-process, producing solar system like abundances, is that $Y_e$ be less then 1/2.  If $Y_e$ is greater then 1/2, all of the neutrons may be quickly incorporated into alpha particles leaving only free protons.

In a core collapse supernova, high neutrino luminosities eject some baryons from the surface of the protoneutron star into a neutrino driven wind.  Many authors have explored r-process nucleosynthesis in this wind \cite{woo94,tak94,qia96}. A number of detailed recent simulations have had difficulties reproducing a successful r-process in this wind, see below.  In this paper we express some of these difficulties in a more model independent form by using data from Supernova Sn1987A and plausible physical assumptions to constrain the r-process.  It is hoped this model independence will help clarify assumptions and problems in the search for the r-process site.

The electron fraction $Y_e$ in the wind is set by the relative rates of the neutrino capture reactions,
\begin{equation}
\nu_e + n \rightarrow p + e^-,
\label{1}
\end{equation}
\begin{equation}
{\bar\nu_e} + p \rightarrow n + e^+.
\label{2}
\end{equation}
These rates depend on the known cross sections and the neutrino and antineutrino luminosities and mean energies.  The cross section for Eq. (\ref{1}) is larger than that for Eq. (\ref{2}) because of important weak magnetism and recoil corrections.  

The ratio of rates for Eqs. (\ref{1}) and (\ref{2}) yields the initial electron fraction $Y_e$ \cite{hor99},
\begin{equation}
Y_e=\left[1+{\bar L \bar\epsilon\over L \epsilon}Q(\epsilon,\bar\epsilon)C(\epsilon,\bar\epsilon)\right]^{-1}.
\label{3}
\end{equation}
Note as nuclei form, $Y_e$ will differ from Eq. (\ref{3}).  However, if this initial $Y_e$ is not significantly neutron rich, see below, a successful r-process is unlikely.  In Eq. (\ref{3}) $\bar L$ is the $\bar\nu_e$ luminosity, $L$ the $\nu_e$ luminosity and the mean $\nu_e$ energy is,
\begin{equation}
\epsilon = \langle E_{\nu_e}^2 \rangle / 
\langle E_{\nu_e} \rangle ,
\label{4}
\end{equation}
where the angle brackets indicate an average over the neutrino spectrum.  We chose this definition for $\epsilon$ because the lowest order cross section is proportional to $\langle E_{\nu_e}^2 \rangle$ while the luminosity includes $\langle E_{\nu_e} \rangle$.  We note, for a Boltzmann spectrum of temperature $T$, $\epsilon= 4T=4/3 \langle E \rangle$ while $\langle E \rangle=3T$. Likewise the mean $\bar\nu_e$ energy is,
\begin{equation}
\bar\epsilon=\langle E_{\bar\nu_e}^2 \rangle/\langle E_{\bar\nu_e} \rangle.
\label{5}
\end{equation}

The neutron proton mass difference $\Delta=M_n-M_p=1.29$\ MeV or reaction $Q$ value contributes the factor $Q(\epsilon,\bar\epsilon)$,
\begin{equation}
Q(\epsilon,\bar\epsilon)={1-2{\Delta\over \bar\epsilon} + a_0 {\Delta^2\over\bar\epsilon^2}\over
1+ 2{\Delta\over \epsilon} + a_0 {\Delta^2\over\epsilon^2}}\, .
\label{6}
\end{equation}
The recoil and weak magnetism corrections to the cross section contribute the factor $C(\epsilon,\bar\epsilon)$,
\begin{equation}
C(\epsilon,\bar\epsilon)={1-7.22 a_2 {\bar\epsilon\over M}\over 1+1.02 a_2{\epsilon\over M}}\, .
\label{7}
\end{equation}
Here $M$ is the nucleon mass and,
$a_0={\langle E_{\nu_e}^2 \rangle / \langle E_{\nu_e} \rangle^2}$,
$a_2= { \langle E_{\nu_e}^3 \rangle \langle E_{\nu_e} \rangle /\langle E_{\nu_e}^2 \rangle^2}$,
describe spectral shapes.  For simplicity we assume $a_0\approx a_2\approx 1.2$ for both neutrinos and antineutrinos \cite{hor99}.  Although the factor $C$ can increase $Y_e$ by 20\%, it is neglected in many supernova simulations.

Equation (\ref{3}) involves observable quantities, the neutrino luminosities $L$, $\bar L$ and mean energies $\epsilon$ and $\bar\epsilon$.  Therefore, one should be able to deduce $Y_e$ from observations of the next galactic supernova.  This is a striking feature of r-process nucleosynthesis in neutrino driven winds.  Perhaps the most important parameter $Y_e$ can be directly probed with observations.  If observations suggest that $Y_e$ is not suitable for the r-process (see below), then the neutrino driven wind may be strongly disfavored as an r-process site.

Supernova SN1987A already provides important data for Eq. (\ref{3}).  Jegerlehner, Neubig and Raffelt \cite{jeg96} place limits on the time averaged antineutrino temperature $T_{\bar\nu_e}$ from Kamikande and IMB observations. We discuss time dependence below.   If one neglects neutrino oscillations, Jegerlehner et al's 95\% upper limit is,
\begin{equation}
T_{\bar\nu_e} < 4.6\ {\rm MeV},
\label{9}
\end{equation}
assuming a Maxwell-Boltzmann spectrum for which,
\begin{equation}
\bar\epsilon=4 T_{\bar\nu_e} < 18.4\ {\rm MeV}.
\label{10}
\end{equation}

If one assumes neutrino oscillations with parameters of the large mixing angle solution to the solar neutrino problem and if one assumes the muon or tau anti-neutrino temperatures are 1.7 times the electron anti-neutrino temperature then the limit decreases to \cite{jeg96},
\begin{equation}
T_{\bar\nu_e} < 4.2\ {\rm MeV},
\label{11}
\end{equation}
\begin{equation}
\bar\epsilon=4T_{\bar\nu_e} < 16.8 \ {\rm MeV}.
\label{12}
\end{equation}
Neutrino oscillations occur after the neutrinos pass through the wind and mix some hot $\bar\nu_\mu$ or $\bar\nu_\tau$ into $\bar\nu_e$.  Therefore the original $\bar\nu_e$ spectrum must have been even colder so that the mixed spectrum could be consistent with observation.  Note, Eq. (\ref{12}) would be even lower but for matter effects as SN1987A neutrinos passed through the Earth.  

Originally, some groups assumed theoretical neutrino spectra and tried to use SN1987A data to set limits on oscillation parameters.  Here, we assume that the oscillation parameters will be determined by independent (solar neutrino) experiments.  Then, one should use the most accurate oscillation parameters to infer neutrino spectra from SN1987A data.  We note that oscillations do not change our results very much.  However, we include both oscillation and non oscillation cases for completeness.     

The ratio of anti-neutrino to neutrino luminosities is constrained by lepton number conservation.  We assume,
\begin{equation}
{\bar L\over L} \leq 1.1\, .
\label{13}
\end{equation}    
Oscillations among active $\nu_e$, $\nu_\mu$, $\nu_\tau$, flavors are not expected to greatly change this.  However oscillations of electron neutrinos to sterile neutrinos (without charged current interactions) $\nu_e\rightarrow \nu_s$ could increase $\bar L/L$ in the wind \cite{nun97,mcl99,cal00}. 

Figure 1 shows $Y_e$ contours for different $\epsilon$ and $\bar\epsilon$ assuming $\bar L=1.1 L$.  The wind is neutron rich in the region to the upper left of the $Y_e=0.5$ contour.  The mean anti-neutrino energy $\bar\epsilon$ must lie within the shaded region, assuming no oscillations, or between the dot-dashed lines assuming oscillations.  This sets a maximum value for $\epsilon$.  In order for the wind to be neutron rich,
\begin{equation}
\epsilon < 11.6\ {\rm MeV},\ \ \ T_{\nu_e} ={\epsilon\over 4} < 2.9 \ {\rm MeV},
\label{14}
\end{equation}
without oscillations and,
\begin{equation}
\epsilon < 10.3\ {\rm MeV},\ \ \ T_{\nu_e} ={\epsilon\over 4} < 2.6 \ {\rm MeV},
\label{15}
\end{equation}
including oscillations.

The requirement $Y_e<0.5$ is a reasonable minimum for the r-process.  If one requires that the wind be significantly neutron rich, for example $Y_e<0.4$ than the limits become,
\begin{equation}
\epsilon < 6.7\ {\rm MeV},\ \ \ T_{\nu_e} ={\epsilon\over 4} < 1.7 \ {\rm MeV},
\label{16}
\end{equation}
without oscillations and,
\begin{equation}
\epsilon < 5.9\ {\rm MeV},\ \ \ T_{\nu_e} ={\epsilon\over 4} < 1.5 \ {\rm MeV},
\label{17}
\end{equation}
including oscillations.  {\it The limits in Eqs. (\ref{14}) through (\ref{17}) are significantly colder then most supernova simulations.}

Hoffman et al. \cite{hof97} discusses a range of physical conditions for the r-process.  We consider two examples.  The first scenario assumes a very short expansion time scale for the wind of order milliseconds \cite{tho01,car97}.  This scenario can proceed with a relatively high $Y_e\approx 0.48$ because the number of seed nuclei formed is reduced by the short time scale.  As a result, the ratio of free neutrons to seed nuclei can still be large enough to produce the heaviest elements.

For this scenario, the limits in Eqs. (\ref{14}) and (\ref{15}) are appropriate.  However, the short expansion time scale may require a high neutrino luminosity that only occurs within a short time of core bounce.  There may not be enough time for $Y_e$ to drop significantly in the protoneutron star.  As a result, the opacity for $\bar\nu_e$ may not be drastically different from that for $\nu_e$ and $\epsilon$ may not be much smaller than $\bar\epsilon$.  Thus it may be difficult to satisfy Eqs. (\ref{14}) or (\ref{15}) at short times.  All realistic supernova simulations that we are aware of, for example \cite{bru91,ram00,bru01}, do not satisfy Eqs. (\ref{14}) and (\ref{15}) at early times ---say within 1/2 second of core bounce. 

A second scenario for the r-process involves a longer expansion time scale for the neutrino driven wind (of order a second).  This can occur at later times in the supernova when the neutrino luminosities are lower.  At later times there can be a large opacity difference between $\bar\nu_e$ and $\nu_e$ so $\epsilon$ can be significantly lower then $\bar\epsilon$.  

However, the longer expansion time scale allows more seed nuclei to form.  Therefore, one will need more free neutrons to have an acceptable ratio of neutrons to seeds.  Thus $Y_e$ must be smaller, for example of order $Y_e<0.4$ (assuming an entropy per baryon of order 130).  The stringent limits in Eqs. (\ref{16}) and (\ref{17}) may be appropriate for this longer expansion time scale r-process.  These limits appear to be very hard to meet.  Simulations tend to give much higher $\nu_e$ temperatures.  Again, we are not aware of any realistic simulation that satisfies them.

We now discuss the time dependence of $\bar\epsilon$.  The cooling of the protoneutron star suggests that $\bar\epsilon$ should decrease with time.  However, as the star becomes more neutron rich, the antineutrino sphere moves in, to higher densities and temperatures.  This could cause $\bar\epsilon$ to increase with time.  The SN1987A data suggest $\bar\epsilon$ {\it decreases} with time. The average electron energy for the six events detected at IMB in the first five seconds is 36 MeV while the remaining two events at later times had an average energy of 20 MeV \cite{IMB}.  Likewise, the average electron energy of the eight events detected at Kamikande during the first five seconds is 17 MeV while the remaining three events had an average energy of 10 MeV \cite{KII}.  Note, some of the difference in energy for the two detectors is explained by IMB's higher threshold.  Clearly the data do not show a rapid {\it increase} of $\bar\epsilon$ at late time.  However the statistics are poor and we have not reanalyzed the data to set quantitative limits on the time dependence.

If the r-process occurs at early times, as might be necessary to have a very short expansion time scale, then the time dependence is not an issue.  The bulk of the SN1987A events will directly constrain $\bar\epsilon$.  However, if the r-process occurs at late times, as is often assumed, then the time dependence could be more important.  We emphasize that a very rapid increase of $\bar\epsilon$ with time, at late times, could evade our limits.  However, there is no positive evidence from SN1987A for this rapid time dependence.  

Of course, one can evade our limits by assuming the r-process occurs in events with different neutrino spectra from those in SN1987A.  This allows one to consider larger $\bar\epsilon$.  However, one may still have to explain a large ratio of $\bar\epsilon$ to $\epsilon$ and why the events are different from SN1987A.  Some alternative sites for the r-process include, prompt supernova explosions \cite{sum01}, 'peculiar' supernovae \cite{wan01}, jets \cite{nag01} and colliding neutron stars \cite{jan01,lat74,ros00}.

Equation (\ref{3}) includes weak magnetism and recoil corrections to the cross sections that are neglected in most supernova simulations.  These are clearly important for $Y_e$ in the wind.  Therefore, they should be included in simulations.   The reduction in the $\bar\nu_e$ cross section from weak magnetism may slightly raise the $\bar\epsilon$ or $\bar L$ predicted by the simulations.  However, $\bar\epsilon$ is constrained by SN1987A data while $\bar L/L$ is constrained by lepton number conservation.  Simulations must still satisfy our limits on $\epsilon$ independent of their inclusion of weak magnetism and recoil corrections.    

We use the Jegerlehner et al. results for $\bar\epsilon$ from SN1987A because of their simplicity. These results could depend somewhat on assumed spectral shapes, or other details.  For example Janka and Hillebrandt \cite{jan89} analyze SN1987A data assuming a Fermi Dirac spectrum with degeneracy $\eta$, $f(E)=[1+{\rm e}^{(E/T-\eta)}]^{-1}$.  They find $\eta=0$ is, somewhat weakly, favored with an upper bound of 2.5 and $T_{\bar\nu_e}<4.5$ MeV.  For $\eta=0$ this is in good agreement with Eq. (\ref{10}) while $\eta=2.5$ increases our limit in Eq. (\ref{10}) by 13\% to $\bar\epsilon<20.8$ MeV.   This, in turn, allows a 13\% increase in the mean neutrino energy in Eq. (\ref{14}) to $\epsilon<13.2$ MeV.  However, if the $\nu_e$ spectrum has the same $\eta$ as the antineutrino spectrum, then the limit on the $\nu_e$ temperature in Eq. (\ref{14}) is unchanged $T_{\nu_e}<2.9$ MeV.  Thus, one can only increase the limit on $T_{\nu_e}$ by assuming a smaller $\eta$ for $\nu_e$ then for $\bar\nu_e$.  Note, we have not included small corrections from changes in $a_0$ and $a_2$.  The dependence of our bounds on spectral shape will be discussed further in future work.

In conclusion:\\
1) The electron fraction in the wind depends on observable neutrino luminosities and mean energies.  Observations of the next galactic supernova should determine $Y_e$.
\\
2) The mean antineutrino energy $\bar\epsilon$ is already limited by SN1987A data while the ratio of antineutrino to neutrino luminosities is constrained by lepton number conservation.  
\\
3) To obtain a $Y_e$ suitable for an r-process in the neutrino driven wind, the mean electron neutrino energy could be significantly lower then that in present simulations.  Alternatively, there could be a rapid increase of the mean antineutrino energy at late times.  However, this is not seen in SN1987A data.  Finally, there could be new neutrino physics such as oscillations to sterile neutrinos.

\acknowledgments

We thank John Beacom, Hans-Thomas Janka, Bradley Meyer, Mitesh Pitel, Georg Raffelt and Todd Tompson for many useful discussions and comments on the manuscript. We thank the Institute for Nuclear Theory for its hospitality during much of this work and acknowledge financial support from DOE grant DE-FG02-87ER40365.

\clearpage
\begin{figure}[h]
\leavevmode\centering\psfig{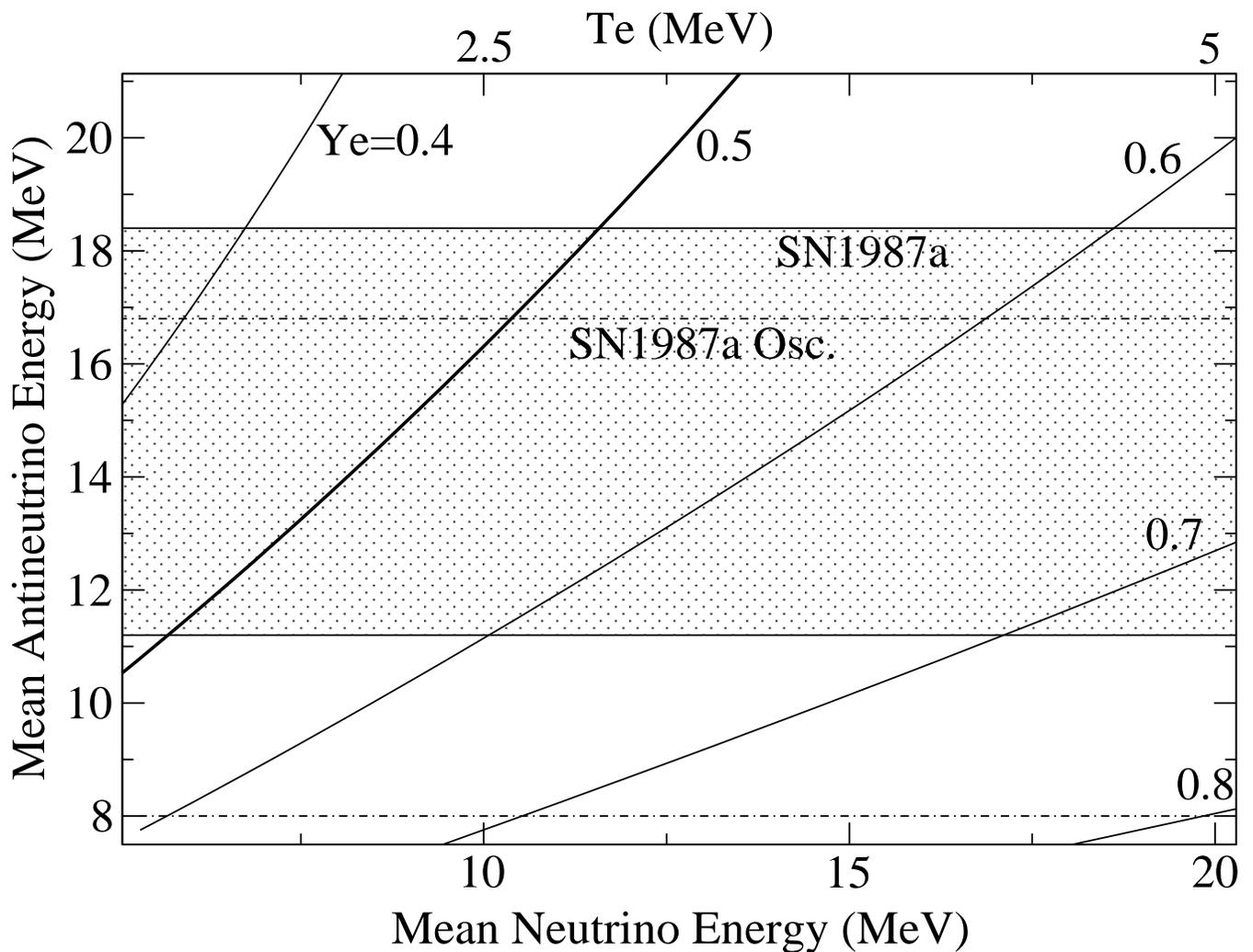}
\caption{Mean electron anti-neutrino energy $\bar\epsilon=\langle E_{\bar\nu_e}^2 \rangle / \langle E_{\bar\nu_e} \rangle$ versus mean electron neutrino energy $\epsilon=\langle E_{\nu_e}^2 \rangle / \langle E_{\nu_e} \rangle$.  Contours of constant electron fraction are indicated for $Y_e$ values from 0.4 to 0.8. For a Boltzmann spectrum the electron neutrino temperature is $1/4$ of $\epsilon$ as indicated by the upper x-axis scale.  The mean antineutrino energy $\bar\epsilon$ must be in the shaded region to be consistent with SN1987A data assuming no neutrino oscillations and between the dot dashed lines assuming oscillations with the large mixing angle MSW solar neutrino masses and mixing angles.
}
 \label{Fig1}
\end{figure}

\clearpage 

\end{document}